\documentclass[preprintnumbers,prd,twocolumn,nofootinbib,
superscriptaddress,nobibnotes,showpacs]{revtex4}
\usepackage{amsfonts}
\usepackage{mathrsfs}
\usepackage{epsfig}
\usepackage{graphicx}
\usepackage{dcolumn}
\usepackage{amsmath}
\usepackage{color}

\def\be{\begin{equation}}
\def\ee{\end{equation}}
\def\bea{\begin{eqnarray}}
\def\eea{\end{eqnarray}}
\def\fnl{f_{\rm NL}}

\def\gnl{g_{\rm NL}}

\begin{document}

\title{Primordial black holes as a tool for constraining non-Gaussianity}


%\author{}
%\email[]{}
%\thanks{}
%\affiliation{}
\author{Christian T. Byrnes} 
\email[]{cbyrnes@cern.ch}
\affiliation{CERN, PH-TH Division, CH-1211, Geneva 23, Switzerland.} 
\author{Edmund J. Copeland} 
\email[]{ed.copeland@nottingham.ac.uk}
\author{Anne M. Green}
\email[]{anne.green@nottingham.ac.uk}
\affiliation{School of Physics and Astronomy, University of 
Nottingham, University Park, Nottingham, NG7 2RD, UK}

%\date{\today}

\begin{abstract}
Primordial Black Holes (PBH's) can form in the early Universe from the collapse of large density fluctuations. Tight observational limits on their abundance constrain the amplitude of the primordial fluctuations on very small scales which can not otherwise be constrained, with PBH's only forming from the extremely rare large fluctuations. The number of PBH's formed is therefore sensitive to small changes in the shape of the tail of the fluctuation  distribution, which itself depends on the amount of non-Gaussianity present. We study, for the first time, how quadratic and cubic local non-Gaussianity of arbitrary size (parameterised by  $\fnl$ and $\gnl$ respectively)
affects the PBH abundance and the resulting constraints on the amplitude of the fluctuations on very small scales.
Intriguingly we find that even non-linearity parameters of order unity have a significant impact on the PBH abundance. The sign of the non-Gaussianity is particularly important, with the constraint on the allowed fluctuation amplitude tightening by an order of magnitude as $\fnl$ changes from just $-0.5$ to $0.5$. We find that if PBH's are observed in the future, then regardless of the amplitude of the fluctuations, non-negligible negative $\fnl$ would be ruled out. Finally we show that $\gnl$ can have an even larger effect on the number of PBH's formed than $\fnl$.
%hence PBHs can be thought of as excellent tools for constraining acceptable levels of non-Gaussianity}. 
\end{abstract}

\pacs{95.35.+d \hfill CERN-PH-TH/2012-167}
%\keywords{}

\maketitle

%\section{}
\section{Introduction}

Primordial Black Holes (PBH's) play a very special role in cosmology. They have never been detected but this very fact is enough to rule out or at least tightly constrain many cosmological paradigms. Convincing theoretical arguments suggest that during radiation domination they can form from the collapse of large density fluctuations~\cite{ch}. If the density perturbation at horizon entry in a given region exceeds a threshold
value, of order unity, then gravity overcomes pressure forces and the
region collapses to form a PBH with mass of order the horizon mass.

There are tight constraints on the abundance of PBH's formed due to their gravitational effects
and the consequences of their evaporation (for recent updates and compilations of the constraints see Refs.~\cite{Josan:2009qn,Carr:2009jm}).  These abundance
constraints can be used to constrain the primordial power spectrum,
and hence models of inflation, on scales far smaller than those
probed by cosmological observations (e.g.~Refs.~\cite{Carr:1994ar,Green:1997sz, Peiris:2008be,Josan:2010cj}). These calculations usually assume that the primordial fluctuations are Gaussian. However since PBH's form from the extremely rare, large fluctuations in the tail of the fluctuation distribution non-Gaussianity can potentially significantly affect the number of PBH's formed. Therefore PBH formation probes both the amplitude and the non-Gaussianity of the primordial fluctuations on small scales. 

Bullock and Primack~\cite{Bullock:1996at} and Ivanov~\cite{Ivanov:1997ia} were the first to study the effects of non-Gaussianity on PBH formation, reaching opposite conclusions on whether non-Gaussianity 
enhances or suppresses the number of PBH's formed (see also Ref.~\cite{Yokoyama:1998xd}).  Refs.~\cite{Hidalgo:2007vk,Saito:2008em} used a non-Gaussian probability distribution function (pdf) derived from an expansion
about the Gaussian pdf~\cite{Seery:2006wk,LoVerde:2007ri} to study PBH formation. However, since PBH's form from rare fluctuations in the extreme tails of the probability distribution, expansions which are only valid for typical size fluctuations can not reliably be used to study PBH formation.

Ref.~\cite{Lyth:2012yp} studied the constraints from PBH formation on the primordial curvature perturbation
for the special cases where it has the form $\zeta =  \pm (x^2- \langle x^2 \rangle)$, where $x$ has a Gaussian distribution, see also Ref.~\cite{PinaAvelino:2005rm}. The minus sign is expected from the linear era of the hybrid inflation waterfall (see also Ref.~\cite{Bugaev:2011wy}), while the positive sign might arise if $\zeta$ is generated after inflation by a curvaton-type mechanism.  

In this paper we go beyond this earlier work and calculate the constraints on the amplitude of the primordial curvature fluctuations, $\zeta$, from black hole formation for both the quadratic and cubic local non-Gaussianity models
(parameterised by $\fnl$ and $\gnl$ respectively). In the process we calculate the probability distribution function of the curvature perturbation for these models. Our results are valid for arbitrary values of these non-linearity parameters, and we recover the known limiting results for very small or large non-Gaussianity. In Sec.~\ref{sec-PBH} we review the calculation of the PBH abundance constraints in the standard case of Gaussian fluctuations, before calculating the constraints for quadratic and cubic local non-Gaussianity in Sec.~\ref{sec:quad} and \ref{sec:cube} respectively. We conclude with discussion in Sec.~\ref{sec:conc}.

\section{Primordial black hole formation constraints}
\label{sec-PBH}

The condition for collapse to form a PBH is traditionally stated in terms of the smoothed density contrast at horizon crossing, $\delta_{\rm hor}(R)$. A fluctuation on a scale $R$ will collapse to form a PBH, with mass $M_{\rm PBH}$ roughly equal to the horizon mass, if $\delta_{\rm hor}(R) > \delta_{\rm c} \sim {\cal O} (1)$~\cite{ch}~\footnote{It was previously thought that there was an upper limit on the size of fluctuations which form PBH's, with larger fluctuations forming a separate closed universe. Kopp et al.~\cite{Kopp:2010sh} have recently shown that this is in fact not the case.}. If the initial perturbations have a Gaussian distribution then the probability distribution of the smoothed density contrast is
given by (e.g. Ref.~\cite{LL}):
\begin{equation}
P(\delta_{\rm hor} (R)) = \frac{1}{\sqrt{2 \pi} \sigma_{\rm hor}(R)}
\exp{ \left( - \frac{\delta_{\rm hor}^2(R)}{2 \sigma_{\rm hor}^2(R)} \right)}\,,
\end{equation}
where $\sigma(R)$ is the mass variance
\begin{equation}
\label{variance}
\sigma^2(R)=\int_{0}^{\infty} \tilde{W}^2(kR)\mathcal{P}_{\delta}(k, t)\frac{{\rm d} k}{k},
\end{equation}
while $\mathcal{P}_{\delta}(k, t)$ is the power spectrum of the
(unsmoothed) density contrast
\begin{equation}
\mathcal{P}_{\delta}(k, t) \equiv \frac{k^3}{2 \pi^2} \langle
|\delta_{k} |^2 \rangle \,,
\end{equation}
and $\tilde{W}(kR)$ is the Fourier transform of the window function used to smooth the density contrast. 

The initial PBH mass fraction
\begin{equation}
\beta(M_{\rm PBH}) \equiv \frac{\rho_{\rm PBH} (M_{\rm PBH})}{\rho_{\rm tot}} \,,
\end{equation}
is equal to the fraction of the energy density of the Universe contained in
regions dense enough to form PBH's which is given by~\footnote{We do not follow the usual Press-Schechter~\cite{Press:1973iz} practice   
of multiplying by a factor of 2 so that all the mass in the Universe is accounted for, since it is not clear if a simple mulitplicative constant is adequate when dealing with non-symmetric configurations.  Furthermore this factor is comparable to other uncertainties in the PBH mass fraction calculation, such as the fraction of the horizon mass which forms a PBH (see e.g. Ref.~\cite{Carr:2009jm}).}
\begin{equation}
\beta(M_{\rm PBH})  =  
      \int_{\delta_{\rm c}}^{\infty} P(\delta_{\rm hor}(R)) \,{\rm d} \delta_{\rm hor}(R) \,.
	               \label{presssch}
\end{equation}
The PBH initial mass fraction is then related to the mass variance by 
\begin{eqnarray}	
\beta(M_{\rm PBH})& =&  \frac{1}{\sqrt{2\pi}\sigma_{\rm hor}(R)} 
\int_{\delta_{\rm c}}^{\infty} \exp{\left(- \frac{\delta^2_{\rm hor}(R)}
    {2 \sigma_{\rm hor}^2(R)}\right)} 
  \,{\rm d}\delta_{\rm hor}(R) \,,\nonumber \\
 &=& \frac{1}{2}  {\rm erfc}\left(\frac{\delta_{\rm c}}{
   \sqrt{2}\sigma_{\rm hor}(R)}\right) \,. 
\label{densitypara}
\end{eqnarray}
The constraints on the PBH initial mass fraction, $\beta(M_{\rm PBH})$, can therefore be
translated into constraints on the mass variance by inverting this expression. There are a wide range of  constraints on the PBH abundance, from their various gravitational effects and the consequences of their evaporation, which apply over different mass ranges. These constraints are mass dependent and lie in the range $\beta(M_{\rm PBH}) < 10^{-20} - 10^{-5}$~\cite{Josan:2009qn,Carr:2009jm}. The power of these PBH abundance constraints is apparent when we consider the 
resulting constraints on $\sigma_{\rm hor}(R)$ which are in the range $\sigma_{\rm hor}(R)/\delta_{\rm c}  < 0.1-0.2 $. In other words a small change in $\sigma_{\rm hor}(R)/\delta_{\rm c} $ leads to a huge change in $\beta$. 
Finally to impose constraints on the power spectrum of the curvature perturbation the transfer function which relates $\delta(k,t)$ to the primordial curvature perturbation is calculated (e.g. Refs.~\cite{Bringmann:2001yp,Josan:2009qn}).  

For an interesting number of PBH's to form the power spectrum on small scales must be several orders of magnitude larger than on a cosmological scales.
This is possible in models, such as the running mass model~\cite{Stewart:1996ey,Stewart:1997wg,Leach:2000ea,Kohri:2007qn,Alabidi:2009bk,Drees:2011hb}, where the power spectrum increases monotonically with increasing wave-number~\cite{Peiris:2008be,Josan:2010cj}. Another possibility is a peak in the primordial power spectrum, due to a phase transition during inflation or features in the inflation potential
(see Ref.~\cite{Bugaev:2010bb} and references therein). 

To assess the affects of non-Gaussianity on the PBH constraints on the curvature perturbations it is sufficient to follow Ref.~\cite{Lyth:2012yp} and work directly with the curvature perturbation, with the PBH abundance being given by 
\begin{equation}
\beta =  
      \int_{\zeta_{\rm c}}^{\infty} P(\zeta) \, {\rm d} \zeta \,.
\end{equation}
For Gaussian fluctuations
\begin{equation}\label{Gaussian}
P( \zeta)  = \frac{1}{\sqrt{2 \pi} \sigma}
\exp{ \left( - \frac{\zeta^2}{2 \sigma^2} \right)}\,,
\end{equation}
and hence
\begin{equation}
\label{betagauss}
\beta =  \frac{1}{2} \, {\rm erfc}\left(\frac{\zeta_{\rm c}}{
   \sqrt{2}\sigma}\right) \,, 
\end{equation}
where $\zeta_{\rm c}$ is the threshold for PBH formation. The variance of the probability distribution is related to the power spectrum of the curvature perturbation,
\begin{equation}
\mathcal{P}_{\zeta} \equiv \frac{k^3}{2 \pi^2} \langle
|\zeta_{k} |^2 \rangle \,,
\end{equation}
 by $\sigma^2 \approx {\cal P}_{\zeta}$. Here and subsequently for compactness we drop the explicit scale dependence of $\beta$ and $\sigma$ and the subscript `hor'  indicating that $\sigma$ is to be evaluated at horizon crossing of the mass variance, i.e. $\sigma \equiv \sigma_{\rm hor}(R)$.
 
The constraints on $\sigma$ obtained via this method differ from those obtained from the full calculation involving the smoothed density contrast by ${\cal O}(10 \%)$: for $\beta=10^{-20}$ the full calculation gives ${\cal P}_{\zeta}^{1/2} = 0.12$~\cite{Josan:2009qn}, while using Eq.~(\ref{betagauss}) gives $\sigma ={\cal P}_{\zeta}^{1/2}= 0.11$. We take $\zeta_{\rm c}=1$ for definiteness, however any variation in the threshold for collapse affects the constraints on $\sigma$ in the Gaussian and non-Gaussian cases in the same way (since it is in fact the combination $\sigma/\zeta_{\rm c}$ that is constrained). 
We consider the upper and lower values of the constraints on $\beta$, $10^{-5}$ and $10^{-20}$, and henceforth drop the explicit dependence on the PBH mass.

Because PBH's form from the rare large fluctuations in the tail of the distribution, this corresponds to $\zeta_{\rm c} /(\sqrt{2} \sigma) \gg 1$ which allows a useful analytic approximation for $\sigma(\beta)$ to be found. Using the large $x$ limit of ${\rm erfc}(x)$, Eq.~(\ref{betagauss}) can be rewritten as 
\begin{equation}
\beta \approx \frac{\sigma}{\sqrt{2 \pi} \zeta_{\rm c}} \exp{ \left( - \frac{\zeta_{\rm c}^2}{2 \sigma^2} \right)} \,,
\end{equation}
 and hence
\begin{equation}
\label{sigmagauss}
\frac{\sigma}{ \zeta_{\rm c}} \approx \sqrt{ \frac{1}{\ln{(1/ \beta)}}} \,.
\end{equation}
Up to now we have been concentrating on the large tail limit of a Gaussian distribution. We will now see that the consequence of allowing even a small amount of non-Gaussianity in the distribution can be dramatic in terms of the constraints from PBH formation -- enough to potentially rule out certain values of the non-linearity parameters. We begin in Sec.~\ref{sec:quad} by considering the impact of adding quadratic local non-Gaussianity. 

\section{Quadratic local non-Gaussianity  ($\fnl$)}
\label{sec:quad}

We take the model of local non-Gaussianity to be 
\bea
\label{zeta-fNL} 
\zeta=\zeta_{\rm G}+\frac35 \fnl (\zeta_{\rm G}^2-\sigma^2)\equiv h(\zeta_{\rm G}), 
\eea
where $\zeta_{\rm G}$ is Gaussian with variance $\sigma^2$, i.e. $\langle \zeta_{\rm G}^2\rangle=\sigma^2$. The constant term is subtracted from $\zeta$ such that $\langle\zeta\rangle=0$, as required by the definition of the curvature perturbation. Note that we are not assuming single--field inflation, but the less restrictive assumption that only a single field direction generates $\zeta$, known as single--source inflation \cite{Suyama:2010uj}, and furthermore this degree of freedom might be different from the one which generates the primordial temperature perturbation on the much larger CMB scales, this is for example the case in \cite{Lyth:2012yp}. The variance of $\zeta$ is given by \cite{Byrnes:2007tm}
\bea
 \label{ps}
 {\cal P}_{\zeta}&=&\sigma^2+4 \left( \frac{3 \fnl}{5} \right)^2 \sigma^4 \ln(kL), 
 \eea
where the cut--off scale $L\simeq1/H$ is of order the horizon scale, the second term is a one-loop correction which dominates in the large $\fnl$ limit, $k$ is the scale of interest and $\ln(kL)$ is typically of order unity (see e.g. Ref.~\cite{Lyth:2007jh}). We have implicitly assumed here, for simplicity, that the primordial power spectrum is close to scale-invariant on the scales of interest, as is the case for models where the power spectrum varies monotonically with wavenumber.

The non-Gaussian probability distribution function (pdf) for $\zeta$, $P_{\rm NG} (\zeta)$, can be found by making a formal change of variables 
\begin{equation}
P_{\rm NG}(\zeta) \, {\rm d} \zeta = \sum_{i=1}^{n} \left\lvert  \frac{{\rm d} h_{i}^{-1}(\zeta)}{{\rm d} \zeta}   \right\rvert  P_{\rm G}(h^{-1}_{i}) \, {\rm d} \zeta \,,
\end{equation}
where $P_{\rm G}$ has a Gaussian distribution and the sum is over the $n$ solutions, $h_{i}^{-1}(\zeta)$, of the equation $h(\zeta_G)= \zeta$. 
For quartic local non-Gaussianity  solving Eq.~(\ref{zeta-fNL}) for $\zeta_{\rm G}$ gives two solutions (which we denote with `$\pm$'): 
\bea
\label{h-inverse} 
h^{-1}_{\pm}(\zeta)=\frac{5}{6\fnl}\left[-1\pm\sqrt{1+\frac{12\fnl}{5}\left(\frac{3\fnl\sigma^2}{5}+\zeta\right)}\right], 
\eea
and hence~\cite{Matarrese:2000iz,LoVerde:2007ri}:
\begin{equation}
\label{defpdfzetafnl}
P_{\rm NG} (\zeta) \, {\rm d} \zeta = \frac{{\rm d} \zeta}{\sqrt{2 \pi}\sigma \sqrt{1+\frac{12\fnl}{5}\left(\frac{3\fnl\sigma^2}{5}+\zeta\right)}} \left( \varepsilon_{-} + \varepsilon_{-} \right) \,,
\end{equation}
where 
\begin{equation} \varepsilon_{\pm}=\exp{ \left[ -\frac12\left(\frac{h^{-1}_{\pm}(\zeta)}{\sigma}\right)^2 \right]} \,.  \end{equation}

\begin{figure}
\includegraphics[width=8.5cm]{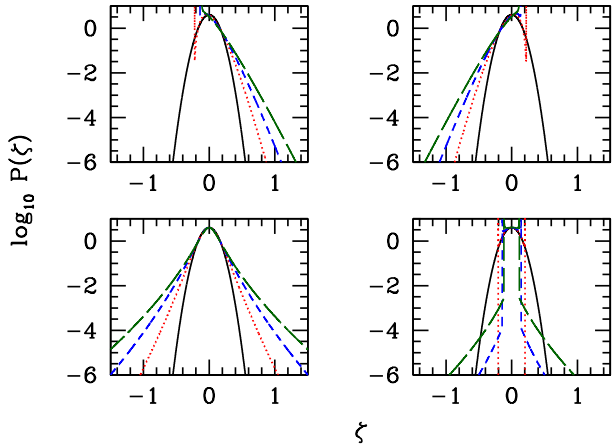}\\
\caption{The log of the non-Gaussian probability distribution, top row $\fnl$ and bottom row $\gnl$.
We have fixed $\sigma=0.1$ throughout and the solid line shows the Gaussian pdf. In the top left panel the dotted, short-dashed and long-dashed lines correspond to $\fnl= 2, 3.5$ and $5$ respectively, while in the top right panel they correspond to $\fnl= -2, -3.5$ and $-5$. In the bottom  left panel the dotted, short-dashed and long-dashed lines correspond to $\gnl= 10, 20$ and $30$ respectively, while in the bottom right panel they correspond to $\gnl= -10, -20$ and $-30$.}
\label{fig:pdfs}
\end{figure}

The log of the non-Gaussian pdf is shown in the upper panels of Fig.~\ref{fig:pdfs} for $\sigma=0.1$ and $\fnl= 0, \pm 2, \pm 3.5$ and $\pm 5$. For positive $\fnl$, as $\fnl$ is increased the amplitude of the large $\zeta$ tail of the pdf increases in amplitude.  There is a minimum value of $\zeta$, $\zeta_{\rm min}= \zeta_{\rm lim}$,
\begin{equation}
\label{zetalim}
\zeta_{\rm lim}= -\frac{5}{12\fnl}\left( 1 + \frac{36\fnl^2\sigma^2}{25}\right) \,,
\end{equation}
at which the pdf diverges, since ${\rm d} h^{-1}(\zeta) / {\rm d} \zeta$ tends to infinity as $\zeta \rightarrow \zeta_{\rm lim}$.
 Initially the peak in the pdf increases in amplitude and moves to negative $\zeta$. As $\fnl$ is increased further the pdf becomes monotonic, increasing continuously with decreasing $\zeta$ down to $\zeta_{\rm min}$. For negative $\fnl$ the pdf is the mirror image of that for positive $\fnl$, and in this case there is a maximum possible value of $\zeta$, $\zeta_{\rm max}=\zeta_{\rm lim}$. 
 
The initial PBH mass fraction is given by
\bea
\beta = I_+ + I_- \,,
\eea
where
\bea 
I_{\pm} \equiv \int^{\zeta_{\rm max}}_{\zeta_{\rm c}} \frac{1}{\sqrt{2 \pi}\sigma \sqrt{1+\frac{12\fnl}{5}\left(\frac{3\fnl\sigma^2}{5}+\zeta\right)}} \varepsilon_\pm {\rm d} \zeta \,.
\eea
The lower limit on the integral is $\zeta_{\rm c} \simeq1$, the threshold value above which a black hole is formed.
The upper limit, $\zeta_{\rm max}$, is the maximum possible value of $\zeta$. For $\fnl>0$, $\zeta_{\rm max}=\infty$, while for $\fnl<0$, $\zeta_{\rm max}= \zeta_{\rm lim}$ as discussed above, with $\zeta_{\rm lim}$ given by Eq.~(\ref{zetalim}). 

\begin{figure}
\includegraphics[width=8.5cm]{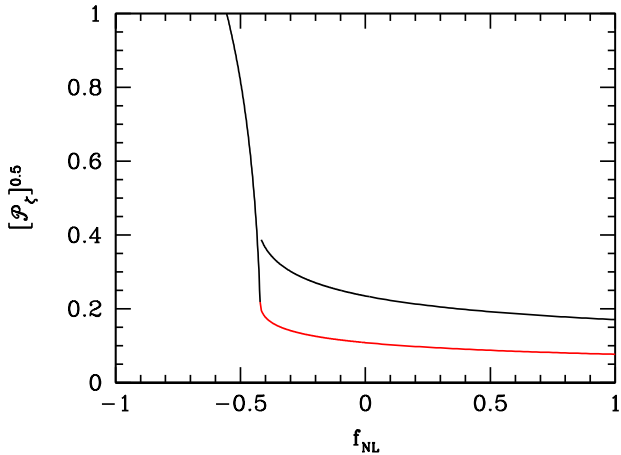}
\caption{The constraints on the square root of the power spectrum of the curvature perturbation, ${\cal P}_{\zeta}^{1/2}$, for initial PBH abundances $\beta=10^{-5}$ and $10^{-20}$ (the upper and lower lines respectively)  for the quadratic local non-Gaussianity model as a function of $\fnl$.}
%for $\fnl < 10$ (upper panel) and $-1 < \fnl < 1$ (bottom panel).}
\label{fig:fNL-small}
\end{figure}

The initial PBH mass fraction is most easily calculated by making a transformation to a new variable $y$,
\bea 
y=\frac{h^{-1}_{\pm}(\zeta)}{\sigma} \,,
\eea
which has unit variance so that
\bea \fnl&>&0:\nonumber \\
\label{betafnlp}
\beta &=& \frac{1}{\sqrt{2\pi}}\left(\int^{\infty}_{y^{\rm c}_+}e^{-y^2/2} \, {\rm d}y +\int^{y^{\rm c}_-}_{-\infty}e^{-y^2/2}\, {\rm d}y\right) \,, \\
& =& \frac{1}{2} {\rm erfc} \left( \frac{y^{\rm c}_+}{\sqrt{2}} \right) + \frac{1}{2} {\rm erfc} \left( \frac{|y^{\rm c}_-|}{\sqrt{2}} \right) \,, \\
\fnl&<&0:  \nonumber \\
\beta &=& \frac{1}{\sqrt{2\pi}}\int^{y^{\rm c}_-}_{y^{\rm c}_+}e^{-y^2/2} \, {\rm d}y  \,, 
 \eea
where $y_{\pm}^{\rm c}$ are the values of $y$ corresponding to the threshold for PBH formation, $\zeta_{\rm c}$:
\bea 
y_{\pm}^{\rm c} =\frac{h^{-1}_{\pm}(\zeta_{\rm c})}{\sigma} \,.
\eea 
For $\fnl>0$,  $y^{\rm c}_+>0$, $y^{\rm c}_-<0$, and $|y^{\rm c}_+|-|y^{\rm c}_-|=y^{\rm c}_+ +y^{\rm c}_-=-3/(5 \fnl)$. Consequently the first integral in the expression  for $\beta$, Eq.~(\ref{betafnlp}), which corresponds to the positive branch, gives the dominant contribution to $\beta$.  However in the limit of very large $\fnl$, $|y^{\rm c}_+| - |y^{\rm c}_-|$ tends to zero and the positive and negative branches contribute equally to $\beta$.

\begin{figure}
\includegraphics[width=8.5cm]{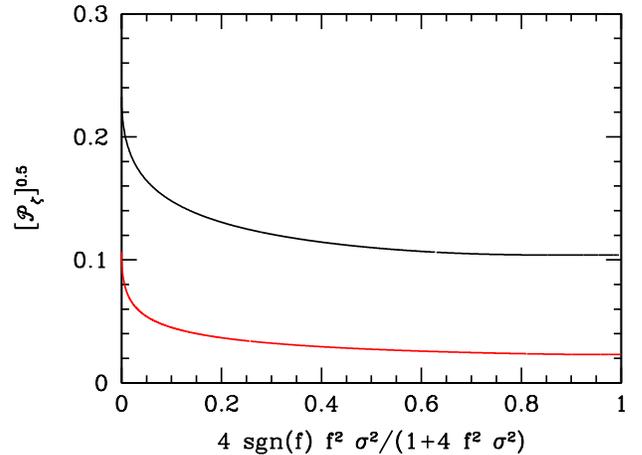}
\caption{The constraints on the square root of the power spectrum of the curvature perturbation, ${\cal P}_{\zeta}^{1/2}$, as in Fig. ~\ref{fig:fNL-small}, for the quadratic local non-Gaussianity model as a function of the fraction of the power spectrum which is non-Gaussian, $4 \, {\rm sgn}(f) f^2 \sigma^2/ (1 + 4 f^2 \sigma^2)$, where $f \equiv 3 \fnl /5$. }
\label{fig:fNL-full-range}
\end{figure}

The constraints on the square root of the power spectrum of the curvature perturbation, ${\cal P}_{\zeta}^{1/2}$, which arise from the tightest and weakest constraints on the initial PBH mass fraction, $\beta < 10^{-20}$ and $10^{-5}$ respectively, are shown in Figs.~\ref{fig:fNL-small} and \ref{fig:fNL-full-range}. In Fig.~\ref{fig:fNL-small} we plot the constraints as a function of $\fnl$ for small $\fnl$, while in Fig.~\ref{fig:fNL-full-range} we plot the constraints as a function of  the fraction of the power spectrum which is non-Gaussian  i.e the ratio of the second non-Gaussian term in the expression for $\langle \zeta^2 \rangle$ in Eq.~(\ref{ps}) to the full expression. The limit that this ratio is unity corresponds to a purely non-Gaussian $\zeta$.  We now discuss how the change in the PBH constraints depends on the amount of non-Gaussianity.

\subsection{Very small $|\fnl|$}

Expanding the pdf for $\zeta$, Eq.~(\ref{defpdfzetafnl}), to second order in $\fnl$ we find
\bea
\label{hs-comparison-pdf}
&&P_{\rm NG}(\zeta) = P_{\rm G}(\zeta) 
   \left[ 1 + \left( \frac{\zeta^2}{\sigma^2} -  3 \right) \frac{
3\fnl\zeta}{5}  \right. \nonumber \\ &&+ \left. \frac{(3\fnl\zeta)^2}{50}\left(\frac{\zeta^4}{\sigma^4}-11\frac{\zeta^2}{\sigma^2}+23-5\frac{\sigma^2}{\zeta^2}\right) \right] \,,
\eea
which agrees with the non-Gaussian pdf in Ref.~\cite{Seery:2006wk} to linear order in $\fnl$.
The second order term has an extra $1/\sigma^2$ term, and hence the above expansion can only be expected to be accurate for $\fnl\ll\sigma^2/\zeta^3$.  Since $\zeta_c\simeq1$ and $\sigma\sim0.1$ in the Gaussian case, this means that this expansion is only valid when applied to PBH formation if $\fnl\lesssim0.01$. 

As can be seen in Fig.~\ref{fig:fNL-small} even a small level of non-Gaussianity has a significant impact on the constraints on $\sigma$. This is because PBH's form from the fluctuations in the extreme tail of the distribution (e.g. for $\sigma\sim0.2$ and $\zeta_c=1$ they are a $5\sigma$ fluctuation) and it is in this regime that even a small skewness is important. The strong asymmetry between positive and negative $\fnl$ is because for $\fnl>0$ overdensities are enhanced, while for $\fnl<0$ the overdensity from a positive $\zeta$ is partially canceled by the Gaussian squared term, thereby $\sigma$ has to become large in order for PBH formation to be possible at all.

\subsection{Intermediate $|\fnl|$}\label{sec:int-fNL}

In the regime where $ 0< \fnl\sigma\ll1$ (which in practice corresponds to $0< \fnl \ll 10$  since $\sigma \sim 0.1$), we have
$\sigma^2 \simeq \langle\zeta^2\rangle$.  However  $\fnl\zeta_{\rm c}\gtrsim1$ and the expression for $h^{-1}(\zeta)$, Eq.~(\ref{h-inverse}), simplifies substantially leading to
\bea
\label{fNL-intermediate} 
\frac{\sigma}{\zeta_{\rm c}}\simeq\left(\frac{5}{3\fnl \zeta_{\rm c}}\right)^{1/2} \sqrt{\frac{1}{2\ln(1/\beta)}} \,.
\eea
The square root term is the result in the Gaussian case, Eq.~(\ref{sigmagauss}). Hence the constraint on ${\mathcal P}_{\zeta}$ is tightened by a factor of $\fnl$ compared to the Gaussian result.

As can be clearly seen from Fig.~\ref{fig:fNL-small}, the constraints are very asymmetric under a change of sign of $\fnl$, with the constraints becoming very rapidly much weaker for $\fnl<0$, we discuss the case of negative $\fnl$ in the next section.

\subsection{Large $|\fnl|$}\label{sec:large-fNL}

In the case of a pure, positive non-Gaussianity the constraints on ${\cal P}_{\zeta}$ become a lot tighter. Since there is a degeneracy between $\fnl$ and $\sigma$ in this case, $\zeta$ can be taken to be given by
\bea 
\zeta=\pm\left(\zeta_{\rm G}^2-\sigma^2\right)=h(\zeta_{\rm G})\leq \sigma^2  \,,
\eea
(c.f. Ref.~\cite{Lyth:2012yp}) and hence $ \langle\zeta^2\rangle\simeq4\sigma^4$.

We first study the $+$ case.
Performing a similar calculation to the more general case with a linear term, the PBH initial mass fraction is given by
\bea 
\beta =\frac{2}{\sqrt{2\pi}}\int_{y_{\rm c}}^{\infty} e^{-y^2/2} {\rm d} y \,,
\eea
and hence~\cite{Lyth:2012yp}
\bea
{\cal P}_{\zeta}^{1/2}\simeq 2\sigma^2\simeq\frac{\zeta_{\rm c}}{\ln(1/\beta)},
 \eea
i.e.~the constraint on $\sigma$ is approximately the square of the constraint in the Gaussian case, and is hence a lot tighter.

The case of a pure, negative chi-squared distribution is very different from the positive case. 
Using the transform $h^{-1}$ this leads to
\bea 
\beta = \int^{\sigma^2}_{\zeta_{\rm c}}  \frac{1}{\sqrt{2 \pi} \sigma\sqrt{\sigma^2-\zeta}}  e^{-(\sigma^2-\zeta)/(2\sigma^2)} \, {\rm d} \zeta \,, 
\eea
which using a further change of variables we transform to
\bea 
\beta = \int^{y_c}_0 \frac{1}{\sqrt{2 \pi y}} e^{-y/2} \, {\rm d} y \,,
\qquad y_c=\frac{\sigma^2-\zeta_{\rm c}}{\sigma^2},  \eea
which leads to the relationship between $\sigma^2\simeq {\cal P}_{\zeta}^{1/2}$ and $\beta$
\bea 
\sigma^2 = \zeta_{\rm c}\left(1+8\pi\sigma^2\beta^2\zeta_{\rm c}+{\cal O}(\beta^4)\right)\simeq\zeta_{\rm c} \,. 
\eea
The constraint on $\sigma$ is very weakly dependent on the limit on $\beta$, confirming the behaviour seen in Fig.~\ref{fig:fNL-small} where the constraints for $\beta=10^{-5}$ and $10^{-20}$ merge for $\fnl \lesssim -0.5$.
%, see also the discussion in Sec.~\ref{sec:int-fNL}. 
For $\fnl < 0$, once $|\fnl| \sigma \gtrsim 0.4$ the pdf for $\zeta$ increases monotonically with increasing $\zeta$, before diverging at $\zeta_{\rm max}=\zeta_{\rm lim}$. If $\zeta_{\rm max} < \zeta_{\rm c}$ then the number of PBH's formed is identically zero, while if $\zeta_{\rm max} > \zeta_{\rm c}$ it is extremely sensitive to the precise value of $\sigma$. Therefore in this regime, unless there is extreme fine-tuning of $\sigma$, the number of PBH's formed will either be completely negligible  or so large that the PBH abundance constraints are violated by many orders of magnitude. Although one can formally produce any given value of $\beta$ with sufficient fine tuning of $\sigma$, in a realistic model the non-Gaussianity will lead to small spatial variations of $\sigma$ in different patches (e.g.~due to a small cubic term ) \cite{Byrnes:2011ri}, which would probably rule this model out if one performed a more detailed calculation. Hence we conclude that a future detection of PBH's would effectively  rule out a negative $\fnl$ unless it has a tiny value, i.e.~from Fig.~\ref{fig:fNL-small}, $\fnl\lesssim-0.5$ would be ruled out regardless of the value of $\sigma$.

Having looked at the case of adding a quadratic type of local non-Gaussianity, we now consider the case of adding a cubic type to see what new constraints this may impose.

\section{Cubic non-Gaussianity ($\gnl$)}
\label{sec:cube}

The model of local non-Gaussianity with a cubic term (but assuming that $\fnl=0$) is defined by
\bea \label{define-h} \zeta=\zeta_{\rm G}+g\zeta_G^3\equiv h(\zeta_{\rm G}), \qquad g\equiv \frac{9}{25}\gnl, \eea
where we have introduced the definition of $g$ in order to reduce the numerical factors which will appear in many expressions in this section. The variance of $\zeta$ for this model is given by
\bea 
\label{psgnl}
{\cal P}_{\zeta}= \sigma^2\left(1+6g\sigma^2 \ln(kL) +27 g^2 \sigma^4\ln(kL)^2\right), 
\eea
where the second term is a one-loop contribution and the second term a two-loop contribution \cite{Byrnes:2007tm}, which nonetheless dominates in the limit of large $g$, and $\ln(kL)$ is again of order unity. 

For $\gnl>0$ the cubic equation $\zeta=\zeta_{\rm G}+g\zeta_{\rm G}^3\equiv h(\zeta_{\rm G})$ has 
 one real solution for all $\zeta$: 
\bea 
\label{hpgnl}
h^{-1}(\zeta)&=&-\left(\frac{2^{1/3}}{3}\right) \left[ g^2 \left( \zeta+\sqrt{\zeta^2 + \frac{4}{27 g}  } \, \right) \right]^{-1/3} \nonumber \\ 
&+&\frac{1}{2^{1/3} g } \left[ g^2 \left( \zeta+\sqrt{\zeta^2 + \frac{4}{27 g}  } \, \right) \right]^{1/3} \,.
\label{cubic-inverse}
\eea
The PBH initial mass fraction is then given by 
\bea
g&>&0:\nonumber \\
\beta &=& 
 \frac{1}{\sqrt{2\pi}}\int^{\infty}_{y_{\rm c}}e^{-y^2/2} \, {\rm d}y= \frac{1}{2} {\rm erfc} \left( \frac{y_{\rm c}}{\sqrt{2}} \right) \,,
\eea
where $y_{\rm c} = h^{-1}(\zeta_{\rm c})/\sigma$, with $h^{-1}(\zeta_{\rm c})$ given by Eq.~(\ref{hpgnl}).

In the case of negative $\gnl$, the cubic function has a local maximum for positive $\zeta_{\rm G}$, with a peak value at $\zeta= \zeta_{\rm t}$ where
\begin{equation} 
\zeta_{\rm t} \equiv \frac{2}{3 \sqrt{3} \sqrt{-g}} \,.
\end{equation}
Hence the cubic polynomial has three real roots if $|\zeta|< \zeta_{\rm t}$, and otherwise only one real root. 
The transition between one and three real roots for $\zeta=\zeta_{\rm c}$ occurs when $g=g_{\rm t}$ where
\begin{equation}
\label{gt}
g_{\rm t} = -\frac{4}{27 \zeta_{\rm c}^2} \approx -0.15   \,.
\end{equation}
The expression for the PBH initial mass fraction therefore has different forms depending on whether
$g$ is greater or less than $g_{\rm t}$.

If $\zeta_{\rm t} < \zeta_{\rm c}$, equivalently $g < g_{\rm t}$, then 
\begin{eqnarray}
g&<&g_{\rm t}:   \nonumber \\
\beta &=&   \frac{1}{\sqrt{2 \pi}} \int_{-\infty}^{y_{c}} \exp{(-
  y^2/2)} \, {\rm d} y  = \frac{1}{2} {\rm erfc} \left( \frac{|y_{\rm c}|}{\sqrt{2}} \right) \,,
\end{eqnarray}
where $y_{\rm c} = h^{-1}(\zeta_{\rm c})/\sigma$ and 
\bea
\label{gvneg}
h^{-1}(\zeta_{\rm c}) &=& - \frac{2^{1/3}}{3 (-g)^{2/3}} \left[ \zeta_{\rm c} + \sqrt{
    \zeta_{\rm c}^2 - \zeta_{{\rm t}}^2 } \right]^{-1/3} \nonumber \\ && -
\frac{1}{2^{1/3} (-g)^{1/3}} \left[ \zeta_{\rm c} + \sqrt{
    \zeta_{\rm c}^2 - \zeta_{{\rm t}}^2 } \right]^{1/3} \,.
\eea

If $\zeta_t>\zeta_c$, equivalently $g_{\rm t} < g< 0$, there are three roots $h^{-1}_1(\zeta_{\rm c} )<0< h^{-1}_2(\zeta_{\rm c} )    < h^{-1}_3(\zeta_{\rm c} )   $, given by
\begin{eqnarray}
h^{-1}_1(\zeta_{\rm c} )&=& - \frac{2}{\sqrt{3} (-g)^{1/2}} \cos{(\theta/3)} \,, \label{h1} \\
 h^{-1}_2(\zeta_{\rm c} )&=& \frac{1}{\sqrt{3} (-g)^{1/2}} \left[ \cos{(\theta/3)} -
  \sqrt{3}  \sin {(\theta/3)} \right] \,, \\
h^{-1}_3(\zeta_{\rm c} ) &=& \frac{1}{\sqrt{3} (-g)^{1/2}} \left[ \cos{(\theta/3)} +
  \sqrt{3}  \sin {(\theta/3)} \right] \,, \label{h3}
\end{eqnarray}
where
\begin{equation}
\theta = {\rm atan} \left[  \frac{ (\zeta_{{\rm t}}^2 - \zeta_{\rm c}^2)^{1/2}  }{\zeta_{\rm c}}     \right] \,.
\end{equation}
It follows that
\begin{eqnarray}
g_{\rm t} &<&g< 0:   \nonumber  \\
\beta& =& \frac{1}{\sqrt{2 \pi}} \left( \int_{-\infty}^{y_{1}} \exp{(- y^2/2)} \, {\rm d} y \right. \nonumber \\
& &+  \left.
  \int_{y_{2}}^{y_{3}} \exp{(- y^2/2)} \, {\rm d} y \right) \,,  \label{beta-gnl} \\ 
  &=& \frac{1}{2} {\rm erfc} \left( \frac{|y_{1}|}{\sqrt{2}} \right)  +  \frac{1}{\sqrt{2 \pi}} \int_{y_{2}}^{y_{3}} \exp{(- y^2/2)} \, {\rm d} y  \,, \label{beta-gnl2}
\end{eqnarray}
where $y_{i} = h^{-1}_{i}(\zeta_{\rm c})/\sigma$, with $h^{-1}_{ i}(\zeta_{\rm c})$ given by Eqs.~(\ref{h1})-(\ref{h3}).

\begin{figure}
\includegraphics[width=8.5cm]{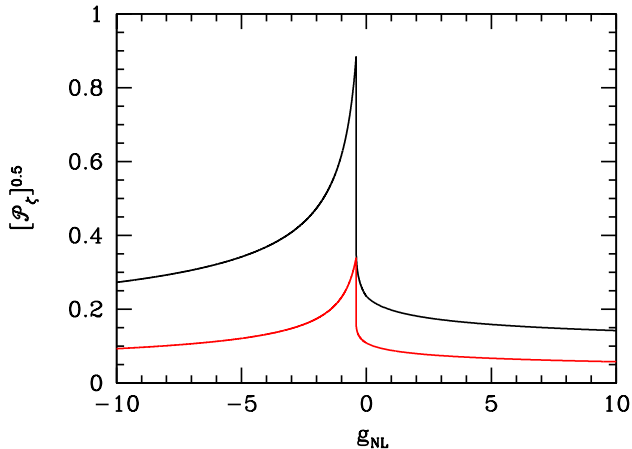}
\caption{The constraints on the square root of the power spectrum of the curvature perturbation, ${\cal P}_{\zeta}^{1/2}$, as in Fig.~\ref{fig:fNL-small}, for the cubic local non-Gaussianity model for $-10< \gnl < 10$.}
\label{fig:gNL-small}
\end{figure}

We calculate the non-Gaussian pdf using the procedure described in Sec. \ref{sec:quad}.
The log of the pdf is shown in the lower panels of Fig.~\ref{fig:pdfs} for $\sigma=0.1$ and $\gnl= 0, \pm 10, \pm 20$ and $\pm 30$. For positive $\gnl$, as $\gnl$ is increased the large $\zeta$ tail of the pdf increases in amplitude, however (unlike the $\fnl$ case) the body of the pdf does not deviate significantly from the Gaussian pdf. This suggests that PBH formation is potentially a more sensitive probe of positive cubic local non-Gaussianity than structure formation.  Negative $\gnl$ is very different from positive $\gnl$. In particular there is a divergence at $\zeta= \zeta_{\rm t}$ which arises from the 
$ {\rm d} h^{-1}(\zeta)/ {\rm d}\zeta$ factor in the pdf. From  Eq.~(\ref{define-h}) we see that 
\be
\frac{{\rm d} h^{-1}(\zeta)}{{\rm d}\zeta} = \frac{1}{1+3g\zeta_G^2} \,.
\ee
This diverges when $1/\zeta_G^2 = 3(-g)$ which from Eq.~(\ref{define-h}) corresponds to $\zeta = \zeta_t$. Expanding around this point with $\zeta = \zeta_t - \delta,~\delta\ll 1$, we find after a little algebra that to leading order 
\be
 \frac{{\rm d} h^{-1}(\zeta)}{{\rm d}\zeta}  = \frac{3^{-1/4}}{2 (-g)^{1/4} \delta^{1/2}} 
 \,.
\ee
This divergence is fairly weak however and the PBH initial mass fraction $\beta$ does not diverge, a result that is confirmed analytically as well as numerically.

\begin{figure}
\includegraphics[width=8.5cm]{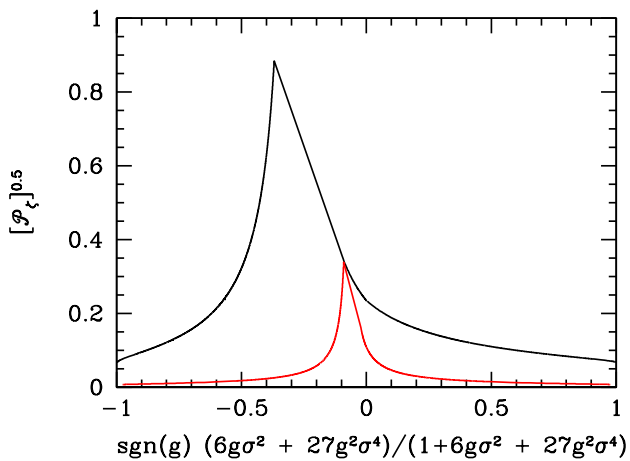}
\caption{The constraints on the square root of the power spectrum of the curvature perturbation, ${\cal P}_{\zeta}^{1/2}$, as in Fig.~\ref{fig:fNL-full-range}, for the cubic local non-Gaussianity model
as a function of the fraction of the power spectrum that is non-Gaussian, ${\rm sgn}(g)(6g\sigma^2+ 27 g^2 \sigma^4)/ (1 +6g\sigma^2+ 27 g^2 \sigma^4)$.}
\label{fig:gNL-full}
\end{figure}

The constraints on the square root of the power spectrum of the curvature perturbation, ${\cal P}_{\zeta}^{1/2}$, which arise from the tightest and weakest constraints on the initial PBH mass fraction, $\beta < 10^{-20}$ and $10^{-5}$ respectively, are shown in Figs.~\ref{fig:gNL-small} and \ref{fig:gNL-full}. In Fig.~\ref{fig:gNL-small} we plot the constraints as a function of $\gnl$ for small $\gnl$, while in Fig.~\ref{fig:gNL-full} we plot the constraints as a function of  the fraction of the power spectrum which is non-Gaussian, i.e.~the ratio of sum of the second and third non-Gaussian terms in Eq.~(\ref{psgnl}) for $\langle \zeta^2 \rangle$ to the full expression.  We now discuss how the change in the PBH constraints depend on the amount of non-Gaussianity.

\subsection{Small $|\gnl|$}

The constraints on the power spectrum are highly asymmetric between positive and negative $\gnl$. This is because for $\gnl>0$ an overdensity in the linear $\zeta$ regime will be boosted by the cubic term, especially strongly in the tail of the distribution and hence the constraint is tightened. However for mildly negative $\gnl$,  the opposite is the case; the two terms tend to cancel each other out and hence the constraints on the power spectrum weaken dramatically in this regime. 
For very small negative $\gnl$ the 2nd term in the expression for $\beta$, Eq.~(\ref{beta-gnl}), the integral of the Gaussian distribution from $y_{2}$ to $y_{3}$ dominates. As  $g \rightarrow g_{\rm t} = -0.15$ from above $y_{3} - y_{2} \rightarrow 0 $ so that this term decreases rapidly and the constraint on the power spectrum rapidly becomes weaker. Only when $g_{\rm t} - g \ll 10^{-80}$ does this term become smaller than the first ${\rm erfc}(|y_{1}|/\sqrt{2})$ term though, and then the value of $\beta$ matches smoothly onto that in the $g<g_{\rm t}$ regime where there is only one real root.  As $g$ is decreased below $g_{\rm t}$ the one real root, $h^{-1}(\zeta_{\rm c})$ given by Eq.~(\ref{gvneg}), becomes less negative and the constraint on
the power spectrum rapidly tightens again.

\subsection{Intermediate $|\gnl|$}

For $\gnl>0$ in the intermediate regime, where the non-Gaussianity is small enough that ${\cal P}_{\zeta}\simeq \sigma^2$, i.e.~$g\sigma^2\ll1$, but large enough to satisfy $g\zeta_{\rm c}^2\gg1$, (hence valid for $1\ll g\ll \sigma^{-2}$),  the expression for $h^{-1}(\zeta_{\rm c})$ in Eq.~(\ref{cubic-inverse}) simplifies significantly to  $h^{-1}(\zeta_{\rm c})  \simeq (\zeta_{\rm c}/g)^{2/3}$. 
Using the leading asymptotic expansion for the Gaussian pdf this leads to
\bea  \frac{\sigma}{\zeta_{\rm c}}&\simeq &\left(\frac{1}{g\zeta_{\rm c}^2}\right)^{1/3} \sqrt{\frac{1}{2\ln(1/\beta)}}. \eea
The term in the square root is the result in the Gaussian case, hence the constraint on $\sigma$ is tightened by a factor of $g^{-1/3}$. For comparison, the result for intermediate quadratic non-Gaussianity is given by Eq.~(\ref{fNL-intermediate}).

To compare the relative size of these changes relative to the Gaussian case, consider a $10 \%$ non-Gaussian correction to $\zeta$, and for concreteness assume that $\sigma=10^{-2}$. Then $g\sigma^2=0.1$ implies $g=10^3$ and the constraint on $\sigma$ tightens by a factor of 10. For a quadratic non-Gaussianity, $3\fnl\sigma/5=0.1$ implies $3\fnl/5=10$ and hence the constraint only tightens by approximately a factor of 3. However, if instead of considering a fixed ratio of non-Gaussian to Gaussian terms, one  considers the non-linearity parameters to have a fixed amplitude then the constraints on $\sigma$ are tightening by the cube root of $g$, but only by the square root of  $f \equiv 3 \fnl/5$.

\subsection{Large $|\gnl|$}
In the limit of very large $|\gnl|$, $\zeta\propto\pm \zeta_{\rm G}^3$ and the constraints don't depend on the sign of $\gnl$. This is because the Gaussian pdf is invariant under a change of sign in $\zeta_{\rm G}$ and this is equivalent to changing the sign of $\gnl$, but only in the case that the linear term is absent. 
As can be seen in Fig.~\ref{fig:gNL-full} the symmetry between positive and negative $\gnl$ only occurs once the modulus of the non-Gaussian fraction of the power spectrum becomes close to one, which corresponds to very large values of $|\gnl|$. In the limit of very large $|\gnl|$ the constraints on ${\cal P}_{\zeta}$ become significantly tighter than in the case of a large and positive $\fnl$. The bound in this case becomes approximately
\bea
 {\cal P}_{\zeta}^{1/2}\simeq 5\sigma^3\simeq \frac{5\zeta_{\rm c}}{(2\ln(1/\beta))^{\frac{3}{2}}} \,, 
 \eea
which is approximately the cube of the bound in the Gaussian case and hence much more stringent (and also tighter compared to the case of a large and positive $\fnl$, as discussed in Sec.~\ref{sec:large-fNL}).

\section{Discussion}\label{sec:conc}

	PBH formation probes the extreme tail of the probability distribution function of the primordial fluctuations. This is the region of the pdf which is most sensitive to the effects of any non-Gaussianity that may be present. We have, for the first time, calculated joint constraints on the amplitude and non-Gaussianity of the primordial perturbations, for arbitrarily large local non-Gaussianity.  We have studied both quadratic and cubic local non-Gaussianity, parameterised by $\fnl$ and $\gnl$ respectively. On the scales associated with the cosmic microwave background and large scale structure, the constraints on primordial non-Gaussianity are approximately $|\fnl|\lesssim10^2$ \cite{Komatsu:2010fb} and $|\gnl|\lesssim10^6$ \cite{Smidt:2010sv}.  In contrast we have shown that on much smaller scales non-linearity parameters of order unity can have a significant effect on the number of PBH's formed. This is because the non-linearity parameters have a larger effect on the tails of the fluctuation distribution than on the more moderate fluctuations probed by cosmological observations. We expect most other forms of non-Gaussianity to also have a significant effect on PBH production, since in general non-Gaussianity generates a skewness which affects the tails of the pdf.

The signs of the non-linearity parameters are particularly important. If positive they always make the constraints tighter by acting in the same direction as the linear contribution to $\zeta$. A negative quadratic term tends to cancel the effect of the linear term, thereby reducing the abundance of large PBH forming fluctuations. The constraints on the amplitude of the power spectrum therefore become much weaker,
of order unity for $\fnl \lesssim -0.5$. In practice this means that the amplitude of fluctuations will either be too small to form any PBH's at all, or so large that almost every horizon region collapses to form a PBH, which is already observationally ruled out. We hence conclude that a future detection of PBH's would rule out a negative value of $\fnl$ unless its value is tiny, $|\fnl|\ll1$.  The case of negative $\gnl$ is different. For $\gnl\simeq-1$ the constraints are weakened as in the negative $\fnl$ case. However as $\gnl$ becomes more negative the constraints quickly become tighter again. In the limit of very large $\gnl$ the constraints are independent of its sign and very tight, approximately the cube of the constraints in the Gaussian case. We have also studied and plotted the probability distribution functions, showing that although the pdfs can diverge, all physical quantities, such as the PBH abundance, remain finite.  The PBH constraints have previously been calculated for a pure $\chi^2$  pdf~\cite{PinaAvelino:2005rm,Lyth:2012yp} and for quadratic non-Gaussianity  in the limits that $|\fnl|\ll1$ and the linear term dominates~\cite{Seery:2006wk,Hidalgo:2007vk}.  We have shown that we recover these limiting cases, however, of particular significance is the fact that our calculations are valid for arbitrarily large quadratic and cubic local non-Gaussianity. 

A bispectrum in the squeezed limit is in general generated by all single field models of inflation, with an amplitude which is related to the spectral index by $\fnl=-5(n_s-1)/12$ \cite{Maldacena:2002vr,Creminelli:2004yq}. Although this value is too small to be seen on CMB scales, it might be important on PBH scales, since firstly the spectral index might be larger as the slow-roll parameters potentially become larger towards the end of inflation and secondly since the constraints are sensitive to smaller values of $\fnl$.

An important issue, which goes beyond the scope of this work, is the calculation of the secondary non-Gaussianities generated through the effects of gravity being non-linear and through horizon re-entry after inflation, during which time $\zeta$ is no longer conserved. These calculations have been carried out on CMB scales and these effects generally cause an order of unity change to the non-linearity parameters (although the effect is scale dependent)~\cite{Bartolo:2010qu,Creminelli:2011sq,Bartolo:2011wb,Junk:2012qt,Lewis:2012tc}. Such small values of the non-linear parameters can have a significant effect on the number of PBH's formed, therefore it would be interesting to carry out a similar analysis valid for the much smaller scales on which PBH's form. If these effects generically lead to $\fnl\sim-1$ then this would suggest that PBH's are unlikely to have formed, unless inflation generated a larger and positive primordial $\fnl$ on the same scales. The current calculation could also be extended by allowing for simultaneous non-zero values of $\fnl$ and $\gnl$ or by studying the effects of higher order non-linearity parameters.

The possibility of PBH formation constrains both the amplitude and degree of non-Gaussianity associated with the primordial density perturbations 
over a wide range of scales, smaller than those probed by cosmological observations. In this paper we have shown that
even relatively small values of the non-linearity parameters, of order unity, can have a significant effect on the PBH constraints on the amplitude of the primordial perturbations. Therefore non-Gaussianity should be taken into account when calculating PBH constraints on inflation models. We have also shown that the observation of PBH's would rule out non-negligible negative $\fnl$, re-emphasizing the constraining power of PBH's.

\begin{acknowledgments}
We would like to thank Peter Coles, Charles Dean-Orange, Shaun Hotchkiss, David Seery and Sam Young for useful discussions. EJC would like to thank the STFC, the
Leverhulme Trust and the Royal Society for financial
support.  AMG  acknowledges support  from STFC. 
 \end{acknowledgments}

\appendix


\begin{thebibliography}{99} 

\bibitem{ch} B.~J.~Carr and S.~W.~Hawking,
Mon.\ Not. \ Roy. \ Astron. Soc. {\bf 168}, (1974) 399; B.~J. Carr {\em Astrophys. J.} {\bf 201} (1975) 1--19. 

\bibitem{Josan:2009qn}
  A.~S.~Josan, A.~M.~Green and K.~A.~Malik,
  %``Generalised constraints on the curvature perturbation from primordial black holes,''
  Phys.\ Rev.\ D {\bf 79} (2009) 103520
  [arXiv:0903.3184 [astro-ph.CO]].
  %%CITATION = ARXIV:0903.3184;%%

\bibitem{Carr:2009jm}
  B.~J.~Carr, K.~Kohri, Y.~Sendouda and J.~Yokoyama,
  %``New cosmological constraints on primordial black holes,''
  Phys.\ Rev.\ D {\bf 81} (2010) 104019
  [arXiv:0912.5297 [astro-ph.CO]].
  %%CITATION = ARXIV:0912.5297;%%

\bibitem{Carr:1994ar}
  B.~J.~Carr, J.~H.~Gilbert and J.~E.~Lidsey,
  %``Black hole relics and inflation: Limits on blue perturbation spectra,''
  Phys.\ Rev.\ D {\bf 50} (1994) 4853
  [astro-ph/9405027].
  %%CITATION = ASTRO-PH/9405027;%%

\bibitem{Green:1997sz}
  A.~M.~Green and A.~R.~Liddle,
  %``Constraints on the density perturbation spectrum from primordial black holes,''
  Phys.\ Rev.\ D {\bf 56} (1997) 6166
  [astro-ph/9704251].
  %%CITATION = ASTRO-PH/9704251;%%

\bibitem{Josan:2010cj}
  A.~S.~Josan and A.~M.~Green,
  %``Constraints from primordial black hole formation at the end of inflation,''
  Phys.\ Rev.\ D {\bf 82} (2010) 047303
  [arXiv:1004.5347 [hep-ph]].
  %%CITATION = ARXIV:1004.5347;%%

\bibitem{Peiris:2008be}
  H.~V.~Peiris and R.~Easther,
  %``Primordial Black Holes, Eternal Inflation, and the Inflationary Parameter Space after WMAP5,''
  JCAP {\bf 0807} (2008) 024
  [arXiv:0805.2154 [astro-ph]].
  %%CITATION = ARXIV:0805.2154;%%

\bibitem{Bullock:1996at}
  J.~S.~Bullock and J.~R.~Primack,
  %``Non-Gaussian fluctuations and primordial black holes from inflation,''
  Phys.\ Rev.\  D {\bf 55} (1997) 7423
  [arXiv:astro-ph/9611106].
  %%CITATION = PHRVA,D55,7423;%%


\bibitem{Ivanov:1997ia}
  P.~Ivanov,
  %``Non-linear metric perturbations and production of primordial black
  %holes,''
  Phys.\ Rev.\  D {\bf 57} (1998) 7145
  [arXiv:astro-ph/9708224].
  %%CITATION = PHRVA,D57,7145;%%

\bibitem{Yokoyama:1998xd}
  J.~'i.~Yokoyama,
  %``Cosmological constraints on primordial black holes produced in the near critical gravitational collapse,''
  Phys.\ Rev.\ D {\bf 58} (1998) 107502
  [gr-qc/9804041].
  %%CITATION = GR-QC/9804041;%%


\bibitem{Hidalgo:2007vk}
  J.~C.~Hidalgo,
  %``The effect of non-Gaussian curvature perturbations on the formation of
  %primordial black holes,''
  arXiv:0708.3875 [astro-ph].
  %%CITATION = ARXIV:0708.3875;%%

\bibitem{Saito:2008em}
  R.~Saito, J.~Yokoyama and R.~Nagata,
  %``Single-field inflation, anomalous enhancement of superhorizon fluctuations,
  %and non-Gaussianity in primordial black hole formation,''
  JCAP {\bf 0806} (2008) 024
  [arXiv:0804.3470 [astro-ph]].
  %%CITATION = JCAPA,0806,024;%%

\bibitem{Seery:2006wk}
  D.~Seery and J.~C.~Hidalgo,
  %``Non-Gaussian corrections to the probability distribution of the  curvature
  %perturbation from inflation,''
  JCAP {\bf 0607} (2006) 008
  [arXiv:astro-ph/0604579].
  %%CITATION = JCAPA,0607,008;%%



\bibitem{LoVerde:2007ri}
  M.~LoVerde, A.~Miller, S.~Shandera and L.~Verde,
  %``Effects of Scale-Dependent Non-Gaussianity on Cosmological Structures,''
  JCAP {\bf 0804} (2008) 014
  [arXiv:0711.4126 [astro-ph]].
  %%CITATION = JCAPA,0804,014;%%

\bibitem{Lyth:2012yp} 
  D.~H.~Lyth,
  %``The hybrid inflation waterfall and the primordial curvature perturbation,''
  JCAP {\bf 1205}, 022 (2012)
  [arXiv:1201.4312 [astro-ph.CO]].
  %%CITATION = ARXIV:1201.4312;%%
  
\bibitem{PinaAvelino:2005rm}
  P.~P. Avelino,
  %``Primordial black hole constraints on non-Gaussian inflation models,''
  Phys.\ Rev.\ D {\bf 72} (2005) 124004
  [astro-ph/0510052].
  %%CITATION = ASTRO-PH/0510052;%%


\bibitem{Bugaev:2011wy}
  E.~Bugaev and P.~Klimai,
  %``Formation of primordial black holes from non-Gaussian perturbations produced in a waterfall transition,''
  Phys.\ Rev.\ D {\bf 85} (2012) 103504
  [arXiv:1112.5601 [astro-ph.CO]].
  %%CITATION = ARXIV:1112.5601;%%


\bibitem{Kopp:2010sh}
  M.~Kopp, S.~Hofmann and J.~Weller,
  %``Separate Universes Do Not Constrain Primordial Black Hole Formation,''
  Phys.\ Rev.\ D {\bf 83} (2011) 124025
  [arXiv:1012.4369 [astro-ph.CO]].
  %%CITATION = ARXIV:1012.4369;%%

\bibitem{LL}
 D.~H. Lyth and A.~R. Liddle, {\em The primordial density perturbation}, Cambridge University Press (2009).

\bibitem{Press:1973iz}
  W.~H.~Press and P.~Schechter,
  %``Formation of galaxies and clusters of galaxies by selfsimilar gravitational
  %condensation,''
  Astrophys.\ J.\  {\bf 187} (1974) 425.
  %%CITATION = ASJOA,187,425;%%

\bibitem{Bringmann:2001yp}
  T.~Bringmann, C.~Kiefer and D.~Polarski,
  %``Primordial black holes from inflationary models with and without broken scale invariance,''
  Phys.\ Rev.\ D {\bf 65} (2002) 024008
  [astro-ph/0109404].
  %%CITATION = ASTRO-PH/0109404;%%

\bibitem{Stewart:1996ey}
  E.~D.~Stewart,
  %``Flattening the inflaton's potential with quantum corrections,''
  Phys.\ Lett.\ B {\bf 391} (1997) 34
  [hep-ph/9606241].
  %%CITATION = HEP-PH/9606241;%%

\bibitem{Stewart:1997wg}
  E.~D.~Stewart,
  %``Flattening the inflaton's potential with quantum corrections. 2.,''
  Phys.\ Rev.\ D {\bf 56} (1997) 2019
  [hep-ph/9703232].
  %%CITATION = HEP-PH/9703232;%%

\bibitem{Leach:2000ea}
  S.~M.~Leach, I.~J.~Grivell and A.~R.~Liddle,
  %``Black hole constraints on the running mass inflation model,''
  Phys.\ Rev.\ D {\bf 62} (2000) 043516
  [astro-ph/0004296].
  %%CITATION = ASTRO-PH/0004296;%%




\bibitem{Kohri:2007qn}
  K.~Kohri, D.~H.~Lyth and A.~Melchiorri,
  %``Black hole formation and slow-roll inflation,''
  JCAP {\bf 0804} (2008) 038
  [arXiv:0711.5006 [hep-ph]].
  %%CITATION = ARXIV:0711.5006;%%

\bibitem{Alabidi:2009bk}
  L.~Alabidi and K.~Kohri,
  %``Generating Primordial Black Holes Via Hilltop-Type Inflation Models,''
  Phys.\ Rev.\ D {\bf 80} (2009) 063511
  [arXiv:0906.1398 [astro-ph.CO]].
  %%CITATION = ARXIV:0906.1398;%%

\bibitem{Drees:2011hb}
  M.~Drees and E.~Erfani,
  %``Running-Mass Inflation Model and Primordial Black Holes,''
  JCAP {\bf 1104} (2011) 005
  [arXiv:1102.2340 [hep-ph]].
  %%CITATION = ARXIV:1102.2340;%%

\bibitem{Bugaev:2010bb}
  E.~Bugaev and P.~Klimai,
  %``Constraints on the induced gravitational wave background from primordial black holes,''
  Phys.\ Rev.\ D {\bf 83} (2011) 083521
  [arXiv:1012.4697 [astro-ph.CO]].
  %%CITATION = ARXIV:1012.4697;%%



%\cite{Suyama:2010uj}
\bibitem{Suyama:2010uj} 
  T.~Suyama, T.~Takahashi, M.~Yamaguchi and S.~Yokoyama,
  %``On Classification of Models of Large Local-Type Non-Gaussianity,''
  JCAP {\bf 1012}, 030 (2010)
  [arXiv:1009.1979 [astro-ph.CO]].
  %%CITATION = ARXIV:1009.1979;%%  

%%%%%%%%%%%%%%%%%%%%%%%%%%%%

%\cite{Byrnes:2007tm}
\bibitem{Byrnes:2007tm} 
  C.~T.~Byrnes, K.~Koyama, M.~Sasaki and D.~Wands,
  %``Diagrammatic approach to non-Gaussianity from inflation,''
  JCAP {\bf 0711}, 027 (2007)
  [arXiv:0705.4096 [hep-th]].
  %%CITATION = ARXIV:0705.4096;%%


%\cite{Lyth:2007jh}
\bibitem{Lyth:2007jh} 
  D.~H.~Lyth,
  %``The curvature perturbation in a box,''
  JCAP {\bf 0712}, 016 (2007)
  [arXiv:0707.0361 [astro-ph]].
  %%CITATION = ARXIV:0707.0361;%%

\bibitem{Matarrese:2000iz}
  S.~Matarrese, L.~Verde and R.~Jimenez,
  %``The Abundance of high-redshift objects as a probe of non-Gaussian initial conditions,''
  Astrophys.\ J.\  {\bf 541} (2000) 10
  [astro-ph/0001366].
  %%CITATION = ASTRO-PH/0001366;%%

%\cite{Byrnes:2011ri}
\bibitem{Byrnes:2011ri} 
  C.~T.~Byrnes, S.~Nurmi, G.~Tasinato and D.~Wands,
  %``Inhomogeneous non-Gaussianity,''
  JCAP {\bf 1203}, 012 (2012)
  [arXiv:1111.2721 [astro-ph.CO]].
  %%CITATION = ARXIV:1111.2721;%%
  


%\cite{Komatsu:2010fb}
\bibitem{Komatsu:2010fb} 
  E.~Komatsu {\it et al.}  [WMAP Collaboration],
  %``Seven-Year Wilkinson Microwave Anisotropy Probe (WMAP) Observations: Cosmological Interpretation,''
  Astrophys.\ J.\ Suppl.\  {\bf 192}, 18 (2011)
  [arXiv:1001.4538 [astro-ph.CO]].
  %%CITATION = ARXIV:1001.4538;%%
  
%\cite{Smidt:2010sv}
\bibitem{Smidt:2010sv} 
  J.~Smidt, A.~Amblard, A.~Cooray, A.~Heavens, D.~Munshi and P.~Serra,
  %``A Measurement of Cubic-Order Primordial Non-Gaussianity (g_{NL} and \tau_{NL}) With WMAP 5-Year Data,''
  arXiv:1001.5026 [astro-ph.CO].
  %%CITATION = ARXIV:1001.5026;%%  


%\cite{Maldacena:2002vr}
\bibitem{Maldacena:2002vr} 
  J.~M.~Maldacena,
  %``Non-Gaussian features of primordial fluctuations in single field inflationary models,''
  JHEP {\bf 0305}, 013 (2003)
  [astro-ph/0210603].
  %%CITATION = ASTRO-PH/0210603;%%
  
%\cite{Creminelli:2004yq}
\bibitem{Creminelli:2004yq} 
  P.~Creminelli and M.~Zaldarriaga,
  %``Single field consistency relation for the 3-point function,''
  JCAP {\bf 0410}, 006 (2004)
  [astro-ph/0407059].
  %%CITATION = ASTRO-PH/0407059;%%  

%\cite{Bartolo:2010qu}
\bibitem{Bartolo:2010qu} 
  N.~Bartolo, S.~Matarrese and A.~Riotto,
  %``Non-Gaussianity and the Cosmic Microwave Background Anisotropies,''
  Adv.\ Astron.\  {\bf 2010}, 157079 (2010)
  [arXiv:1001.3957 [astro-ph.CO]].
  %%CITATION = ARXIV:1001.3957;%%

%\cite{Creminelli:2011sq}
\bibitem{Creminelli:2011sq} 
  P.~Creminelli, C.~Pitrou and F.~Vernizzi,
  %``The CMB bispectrum in the squeezed limit,''
  JCAP {\bf 1111}, 025 (2011)
  [arXiv:1109.1822 [astro-ph.CO]].
  %%CITATION = ARXIV:1109.1822;%%
  
%\cite{Bartolo:2011wb}
\bibitem{Bartolo:2011wb} 
  N.~Bartolo, S.~Matarrese and A.~Riotto,
  %``Non-Gaussianity in the Cosmic Microwave Background Anisotropies at Recombination in the Squeezed limit,''
  JCAP {\bf 1202}, 017 (2012)
  [arXiv:1109.2043 [astro-ph.CO]].
  %%CITATION = ARXIV:1109.2043;%%  

%\cite{Junk:2012qt}
\bibitem{Junk:2012qt} 
  V.~Junk and E.~Komatsu,
  %``Cosmic Microwave Background Bispectrum from the Lensing--Rees-Sciama Correlation Reexamined: Effects of Non-linear Matter Clustering,''
  arXiv:1204.3789 [astro-ph.CO].
  %%CITATION = ARXIV:1204.3789;%%

%\cite{Lewis:2012tc}
\bibitem{Lewis:2012tc} 
  A.~Lewis,
  %``The full squeezed CMB bispectrum from inflation,''
  JCAP {\bf 1206}, 023 (2012)
  [arXiv:1204.5018 [astro-ph.CO]].
  %%CITATION = ARXIV:1204.5018;%%


\end{thebibliography}

\end{document}